# Charge-qubit operation of an isolated double quantum dot

J. Gorman,[1] D. G. Hasko,[1] and D. A. Williams[2]

[1]*Microelectronics Research Centre, University of Cambridge, Cambridge CB3 0HE, United Kingdom*
[2]*Hitachi Cambridge Laboratory, Hitachi Europe Ltd., Cambridge CB3 0HE, United Kingdom*

We have investigated coherent time evolution of pseudo-molecular states of an isolated (leadless) silicon double quantum-dot, where operations are carried out via capacitively-coupled elements. Manipulation is performed by short pulses applied to a nearby gate, and measurement is performed by a single-electron transistor. The electrical isolation of this qubit results in a significantly longer coherence time than previous reports for semiconductor charge qubits realized in artificial molecules.

Quantum computation offers a means for efficiently solving classes of problem that are practically infeasible by conventional computing [1]. One approach to building a solid-state quantum computer is by exploiting quantum states of artificial atoms and molecules realized in quantum-dot systems. The key challenges in producing efficient quantum circuits are to have a system with sufficiently high number of operations within the characteristic coherence time of the qubits, to control the coupling between qubits to form architectures, and to integrate the qubits with manipulation and measurement circuitry. An obvious candidate for performing quantum measurement on a charge qubit is a capacitively-coupled single-electron tunnelling device, such as a quantum point contact [2], or a single-electron transistor [3].

In this Letter, we demonstrate the operation of an isolated double quantum-dot as a charge qubit. All operations (initialization, manipulation, and measurement) are achieved by capacitively-coupled elements only: gates for initialization and manipulation, and single-electron transistor for measurement. This scheme does not require directly-coupled electronic leads to the individual qubits in the architecture. Directly-coupled leads can be prominent sources of decoherence due to the finite tunnel coupling of the reservoirs to the confined electron that defines the qubit. By using only capacitively-coupled gates and eliminating the resistive coupling, the confined electron states that define the qubit are expected to be less susceptible to such sources of decoherence. The scheme also provides increased flexibility in design, since the qubits may be patterned on a two-dimensional lattice in ways which would be impossible with a set of qubits that are attached to leads.

The qubit [see Fig. 1(a)] is embodied in the pseudo-molecular electronic eigenstates of an isolated double quantum-dot (IDQD) [4]. The base material used is an industry-standard silicon-on-insulator wafer with a Phosphorous-doped active region [see Fig. 1(b)], which is 35 nm thick, and provides the confinement in the vertical ($\hat{z}$) direction. The wafer is patterned and etched to form the device elements and the qubit structure, in which the electrons are strongly confined within a double-well potential. The quantum dots that make up the qubit are coupled together by the ~ 20 nm wide constriction, which is depleted of electrons and acts as a tunnel barrier [see Fig. 1(c)]. Similar tunnel barriers are used to couple a single quantum-dot to source and drain leads to form a single-electron transistor. Single-electron devices that make use of such tunnel barriers have been characterized extensively [5, 6], and the electrochemical potentials of the quantum dots can be controlled through externally applied electric fields [7]. It is also possible to modify the tunnel-barrier transmittances by gate voltages. However, these are principally set by the geometrical constrictions between the dots defined in the fabrication of the device.

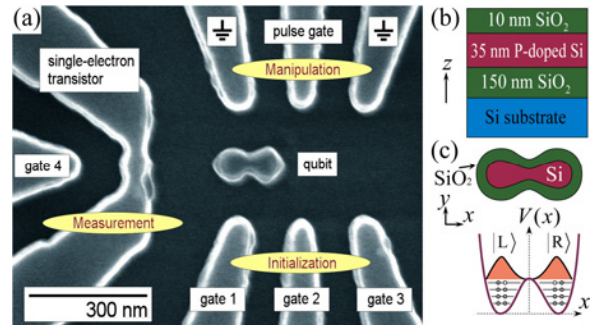

FIG. 1. (a) Scanning electron micrograph of the sample. The device was fabricated by high-resolution electron-beam lithography and reactive-ion etching to 'trench-isolate' the elements. Thermal oxidation was carried out to reduce the dimensions of the silicon regions and to passivate the surface-charge trap states. The device was mounted below the mixing chamber of a dilution refrigerator operating at base temperature (~ 20 mK). (b) Material profile, showing the 35 nm-thick active region between the insulating layers. The dopant (Phosphorous) concentration was $5.7\times10^{19}$ cm$^{-3}$. (c) Schematic view of the IDQD after oxidation, with the active region surrounded by the oxide layer. Localized charge configurations $|L\rangle$ and $|R\rangle$ of the electron at the highest occupation level are illustrated schematically.

The electron at the highest occupation level of the double quantum-dot may be used to embody a charge-qubit as shown in Fig. 2. Because of the absence of resistive coupling, there is no possibility of electrons entering or leaving the double quantum-dot. The two-level quantum system is characterized by the governing energies $\varepsilon$ and $\Delta$, where $\varepsilon$ is the energy difference between the uncoupled charge states $|L\rangle$ and $|R\rangle$ that describe the electron to be fully in the left or the right quantum dot respectively, and $\Delta$ is the anti-crossing energy that arises from the finite interdot coupling when the IDQD tunnel barrier is sufficiently lowered [8]. The energy difference between the coupled stationary eigenstates of the qubit is given by $E^* = E_1 - E_0 = (\varepsilon^2 + \Delta^2)^{1/2}$, illustrated schematically in Fig. 2(a). The device has been extensively characterized in the DC-voltage regime. An operating point (initialization condition) for the qubit is set by the application of stationary electric fields using gates 1, 2 and 3. Manipulation in the time domain [see Fig. 2(b)] is performed by short 'top-hat' pulses on the pulse gate to modify the qubit energies non-adiabatically, and alter the rate of precession of the qubit state vector. Although the manipulation is similar to the method demonstrated for the Cooper-pair box charge qubit [9], the effect of the pulse on the electrically-floating double quantum-dot involves an abrupt change in the self-consistent confinement potential, and modifies $\Delta$ as well as $\varepsilon$. The chosen combination of gate voltages resulted in a typical relaxation time of $\tau_1 \sim 100$ μs to the ground state [10]. The manipulation is repeated in a pulse–train with amplitude $V_P$, duration $\Delta t$, and repetition time $T_r \gg \Delta t$. The pulse-repetition rate was chosen to have sufficiently high measurement signal amplitude, while ensuring that the qubit relaxes to the initialization condition between successive manipulations.

Figure 3(a) shows the measured total current $I$ through the single-electron transistor as a function of $\Delta t$ and $V_{g3}$. The qubit oscillation frequency ω can be changed continuously by the gate voltage $V_{g3}$, and traces the expected nonlinear response as a function of $\varepsilon$. The geometric orientation of the gates dictates that the electric field components from gate 3 are approximately parallel to the longitudinal axis of the double quantum-dot, meaning that a change in $V_{g3}$ should principally influence $\varepsilon$. The maximum oscillation amplitude induced by the qubit on the single-electron transistor is ~5-10% of the average current $I_{max}$ at the conductance maxima from Coulomb blockade, translating to an induced charge which corresponds to $\Delta V_{g4} \approx 7$ mV. This shows a good correlation with the expected order of magnitude of the induced potential difference on the single-electron transistor $\Delta V_{g4}[|L\rangle \to |R\rangle] \sim 10$ mV for the case when a maximum (one-electron) polarization of the double-quantum dot occurs. The inset of Fig. 3(a) shows the measured exponentially-decaying signal amplitude as a function of $T_r$, in the regime $T_r \geq \tau_1 = 100$ μs. Figure 3(b) shows the pulse-amplitude ($V_P$) dependence of the measured current $I$ for a set of experiments at $V_{g2} = 4.5$ V, where $\Delta t$ was swept from 10 ns to 1 μs. Each trace is taken with an incremental increase in pulse amplitude $V_P$ from 100 mV to 500 mV. As expected, the frequency of the exponentially-damped oscillation is seen to increase with $V_P$. Figure 3(c) shows the dependence on the pulse-repetition rate $T_r$, in the regime where $T_r < \tau_1$, to which a dependence of the oscillation frequency is observed. This is attributed to the incomplete relaxation to the ground state between successive operations, and the $T_r$ dependence of the time-averaged charge induced on the qubit, analogous to incrementing a nearby gate voltage. The effect tunes the qubit out of resonance, and also moves the SET away from its charge-sensitive regime. In order to compensate for this effect on the single-electron transistor, we linearly swept $V_{g4}$ simultaneously with $\Delta t$. This maintained the SET near the operating point shown in the inset of Fig. 4(a), and resulted in a clearer oscillation pattern. It should be noted that an exact compensation requires simultaneous sweeping of a multiple number of gates with different proportionality constants, thus making the operation rather cumbersome. However, comprehensive operation of the circuit would involve the single-electron transistor to be effectively off (within the Coulomb gap) during qubit manipulation sequences. The data (dots) fit well to an exponentially damped cosine function (red line) with angular frequency $\omega \approx 62$ MHz, corresponding to a quantum level spacing of $E^* = 40.5$ neV and a coherence time of $\tau_2 = 220$ ns, superimposed on the $\cosh^{-2}$ line–shape of the SET [11]. We also obtained information on the free-evolution dephasing of the qubit by performing a Ramsey-interference experiment, shown in Fig. 4(b), by using $\Delta t_{\pi/2} = 47$ ns obtained from the result shown in Fig. 4(a) [12]. An exponentially-damped sinusoidal oscillation was observed with a coherence time of $\tau_2 \approx 200$ ns, which is slightly lower than the result from the driven experiment, as expected. The signature of an interfering quantum level was also observed (the peak at $\delta = 220$ ns is due to the interference of a secondary sinusoidal oscillation with the main oscillation, with a higher frequency but faster decay). The experiment was limited by the low signal amplitude at the end of the pulse gate because of the attenuation at the silicon part of the coplanar waveguide. This can be overcome by further optimization of the high-frequency setup in the experiment.

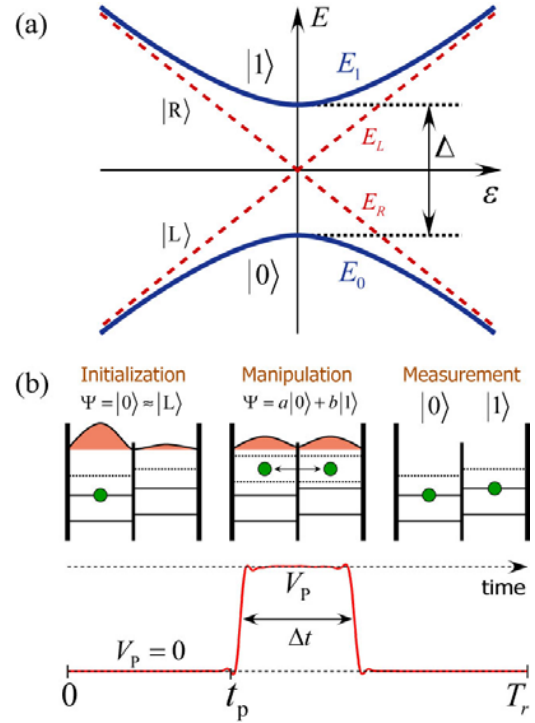

FIG. 2. Scheme for qubit initialization, manipulation, and measurement. (a) Energy diagram illustrating the localized electron states for an uncoupled system $|L\rangle$ and $|R\rangle$ (dashed lines) with eigenenergies $E_L$ and $E_R$, respectively. The inter-dot coupling energy $\Delta$ results in new stationary eigenstates $|0\rangle$ and $|1\rangle$, with eigenenergies $E_0$ and $E_1$, respectively (solid lines). Far from resonance ($|\varepsilon| \gg 0$) the qubit eigenstates are well approximated by $|L\rangle$ and $|R\rangle$, but near resonance ($\varepsilon \approx 0$) the eigenstates are maximally delocalized. (b) Gates 1, 2 and 3 maintain an effective detuning $\varepsilon = E_L - E_R$ and coupling $\Delta$. The pulse is switched on at time $t_p$, and the qubit coherently evolves for the duration $\Delta t$. The lower graph shows the pulse shape for $\Delta t = 5$ ns measured at the output of the pulse generator. At time $t_p + \Delta t$ the pulse is switched off, which stops the manipulation. The resulting quantum superposition determines the SET measurement outcome. The qubit relaxes to the ground state $|0\rangle$ after a time $\sim \tau_1$. The cycle is repeated approximately $10^4$ times. The SET current corresponds to the ensemble average of the measurement outcomes. The SET measurement integration time was 100 ms.

The observed coherence time of 200 ns is two orders of magnitude longer than existing reports for semiconductor double quantum-dot charge qubits [13]. We attribute this result to the weak coupling of the isolated qubit to the noise that results from quantum fluctuations of charge in the surrounding gates, as well as the reduction in the efficiency of the coplanar waveguide in providing a low impedance path for high-frequency noise to travel down to the qubit. Furthermore, the acoustic mismatch of the amorphous oxide, the lack of piezoelectric coupling, and the effect of phonon localization play a part in reducing the electron-phonon interaction on the IDQD structure [14]. Thus, semiconductor charge qubits of this configuration are less susceptible to decoherence in comparison with charge qubits realized through surface-gating schemes.

From our observations, we conclude that the silicon system relies on a non-equilibrium oscillation and a weak interaction with the environment, in a similar manner to liquid-state NMR, allowing operation at a qubit energy splitting less than the ambient thermal energy [15-17]. This explains the long coherence time and the observation of coherent oscillations with a splitting of $E^* \approx 40$ neV at a temperature of $T \geq 20$ mK. However, the small $E^*$ measured in this case means that the number of quantum operations that can be achieved within the coherence time is very limited, and as such the device is far from the $\sim 10^4$ operations that are required of a candidate system to realize fault-tolerant quantum computation. Therefore, an important challenge is to increase the interdot coupling, either by modifying the design of the IDQD structure, or by better gate tuning.

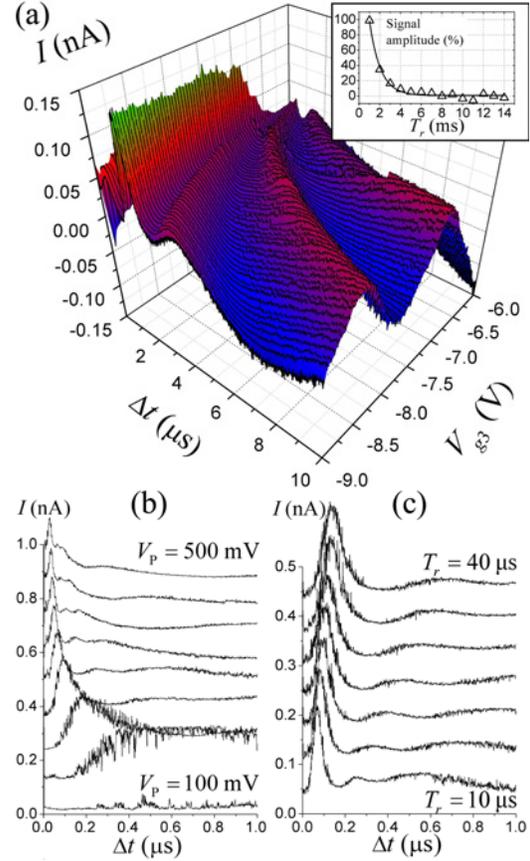

FIG. 3. Measured responses for different experimental parameters. (a) Colour plot of the SET current $I$ as a function of $\Delta t$ and $V_{g3}$, with $V_P = 500$ mV, $T_r = 100$ μs, $V_{g1} = 0$ V, $V_{g2} = -9$ V, $V_{g4} = -0.7$ V, and $V_{sd} = 450$ μV. Measurement of the qubit state is clearly visible. The oscillation frequency $\omega$ can be changed continuously by the gate voltage $V_{g3}$. The pulse also induces a time-averaged (DC) bias, manifested as a small change in $\omega$ as $\Delta t$ is increased. (b) Pulse-height dependence of the coherent oscillations, with $T_r = 10$ μs, $V_{g1} = -6$ V, $V_{g2} = 4.5$ V, $V_{g3} = -6$ V, $V_{g4} = -0.5$ V, and $V_{sd} = 450$ μV. For increasing $V_P$, the oscillation frequency $\omega$ also increases, and the response shifts towards positive effective $V_{g2}$, reducing $\tau_2$. (c) $T_r$ dependence, with $V_P = 300$ mV, $V_{g1} = -6$ V, $V_{g2} = 4.5$ V, $V_{g3} = -6$ V, $V_{g4} = -0.5$ V, and $V_{sd} = 450$ μV. A small dependence on $T_r$ is observed in $\omega$, attributed to the change in the time-averaged induced bias on the qubit for different $T_r$. Data sets in (b) and (c) are offset by 140 pA and 70 pA, respectively.

In conclusion, we have successfully demonstrated charge-qubit operation of an isolated semiconductor artificial molecule, and quantum measurement using a capacitively-coupled single-electron transistor. The principal sources of decoherence are estimated to be the back–action of the 'always-on' SET on the IDQD in our operation scheme, and the electromagnetic noise coupling through the coplanar waveguide. The strength of the measurement implies an equally strong back-action on the IDQD due to the stochastic single-electron tunnelling events through the SET. It should be possible to improve upon the observed coherence time by using an RF-SET [18], where the measurement is switched on and off in synchronism with the manipulation pulse, and the use of more sophisticated filtering in the experimental setup.

This work was supported by Hitachi Europe Ltd. and the LINK-Foresight research project '*Nanoelectronics at the Quantum Edge*' (www.nanotech.org). We thank the members of the LINK project, S. Uno, A. Ferguson, Z. Durrani, H. Qin, R. Collier, and J. Cleaver for useful discussions.

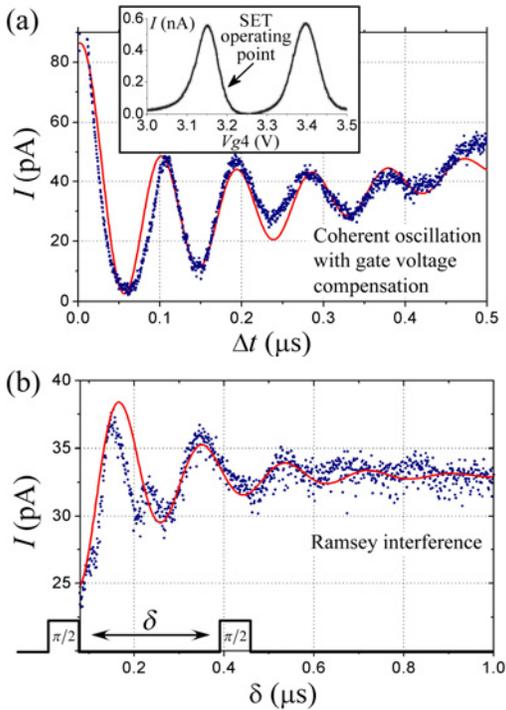

**Fig. 4.** (a) The gate-compensated method, whereby the SET operating point is maintained by proportionally sweeping $V_{g4}$ simultaneously with $\Delta t$. The maximum oscillation amplitude is ~5-10% of the maximum SET current. The inset shows two SET conductance oscillations due to Coulomb blockade, as a function of $V_{g4}$. The arrow indicates the position where the SET was set-up as a highly-sensitive electrometer. (b) Ramsey-interference experiment showing the free-evolution dephasing of the qubit. The experimental parameters were $V_P = 0.6$ V, $\Delta t(\pi/2) = 17$ ns, $V_{g2} = 0.3$ V. The data fits to an exponentially damped sine function (solid line) with a characteristic decay time of $\tau_2 \approx 200$ ns. The small peak at $\delta = 220$ ns is possibly due to a quantum level which is interfering with the qubit evolution. Since the pulse widths are constant, $\varepsilon$, $\Delta$, and $U$ are unchanged, which eliminates the need for compensation by the gates.